\documentclass[12pt]{article}  

\usepackage{cite} 
\usepackage{amssymb} 
\usepackage{amsfonts} 

\normalbaselines
\catcode `\@=11
\@addtoreset{equation}{section}

\def\section{\@startsection {section}{1}{\z@}{-3.5ex plus -1ex minus
     -.2ex}{2.3ex plus .2ex}{\normalsize\bf}}
\def\subsection{\@startsection{subsection}{2}{\z@}{-3.25ex plus -1ex minus
 -.2ex}{1.5ex plus .2ex}{\normalsize\bf}}

\def\thebibliography#1{\section*{References\markboth
  {REFERENCES}{REFERENCES}}\list
  {[\arabic{enumi}]}{\settowidth\labelwidth{[#1]}\leftmargin\labelwidth
  \advance\leftmargin\labelsep
  \usecounter{enumi}}
  \def\newblock{\hskip .11em plus .33em minus -.07em}
  \sloppy
  \sfcode`\.=1000\relax}
 
\catcode `\@=12 

\title{\bf\normalsize PRECANONICAL QUANTIZATION AND
THE SCHR\"ODINGER  WAVE FUNCTIONAL 
}

\author{ 
Igor V. Kanatchikov $^{a,b,}$\thanks{{\sl E-mail}: 
{\tt kai@tpi.uni-jena.de, kai@fuw.edu.pl, 
} 
{\sl URL}: {\tt http://www.tpi.uni-jena.de/$\tilde{\hspace*{3pt}}$kai}.} 
{$^{,}$}\thanks{On leave 
from Tallinn Technical University, 19086 Tallinn, Estonia. } 
\\
$^a$ {\it\small 
Institute of Theoretical Physics, 
Friedrich Schiller University of Jena}\\
{\it\small Max-Wien-Platz 1, D-07743 Jena, Germany}  \\  
$^b$ {\it\small Laboratory of Analytical Mechanics and Field Theory }  \\
 {\it\small Institute of Fundamental Technological Research }  \\
 {\it\small Polish Academy of Sciences, 
Warsaw  00-049, Poland }  
} 

\date{\sf\small 
final version, April 05 2001 \\ 
accepted for publication in Phys. Lett. A} 

\begin{document}

\maketitle




\vspace*{-10mm} 
 \vspace*{-89mm}\vspace*{-2mm} 
\hbox to 6.25truein{
\footnotesize\it 
\hfil \hbox to 0 truecm{\hss 
\normalsize\rm 
{\sf  hep-th/0012084} { }}\vspace*{-3.5mm}}
\hbox to 6.2truein{
\vspace*{-1mm}\footnotesize 
\hfil 
} 
\hbox to 6.25truein{
\hfil \hbox to 0 truecm{ 
\hss \normalsize 
\sf FSU TPI 08/00 
\hfil
}
}

  
\vspace*{65mm} \vspace*{16mm} 




\begin{abstract}
\noindent
A relation between the Schr\"odinger wave functional and the 
Clifford-valued wave function 
which appears in what we call 
precanonical quantization of fields and fulfils a Dirac-like 
 generalized covariant Schr\"odinger 
equation on the space of field and space-time variables  
is discussed. 
The Schr\"odinger wave functional is argued to be 
the trace of the positive frequency part of the 
continual product over all spatial points of the values of the 
aforementioned 
wave function restricted to a Cauchy surface. 
The standard functional differential Schr\"odinger equation 
is derived as a consequence of the Dirac-like 
covariant Schr\"odinger  
equation. 
\vspace*{0.5cm}
\end{abstract} 


\newcommand{\beq}{\begin{equation}}
\newcommand{\eeq}{\end{equation}}
\newcommand{\beqa}{\begin{eqnarray}}
\newcommand{\eeqa}{\end{eqnarray}}
\newcommand{\nn}{\nonumber}

\newcommand{\half}{\frac{1}{2}}

\newcommand{\xt}{\tilde{X}}

\newcommand{\uind}[2]{^{#1_1 \, ... \, #1_{#2}} }
\newcommand{\lind}[2]{_{#1_1 \, ... \, #1_{#2}} }
\newcommand{\com}[2]{[#1,#2]_{-}} 
\newcommand{\acom}[2]{[#1,#2]_{+}} 
\newcommand{\compm}[2]{[#1,#2]_{\pm}}

\newcommand{\lie}[1]{\pounds_{#1}}
\newcommand{\co}{\circ}
\newcommand{\sgn}[1]{(-1)^{#1}}
\newcommand{\lbr}[2]{ [ \hspace*{-1.5pt} [ #1 , #2 ] \hspace*{-1.5pt} ] }
\newcommand{\lbrpm}[2]{ [ \hspace*{-1.5pt} [ #1 , #2 ] \hspace*{-1.5pt}
 ]_{\pm} }
\newcommand{\lbrp}[2]{ [ \hspace*{-1.5pt} [ #1 , #2 ] \hspace*{-1.5pt} ]_+ }
\newcommand{\lbrm}[2]{ [ \hspace*{-1.5pt} [ #1 , #2 ] \hspace*{-1.5pt} ]_- }
\newcommand{\pbr}[2]{ \{ \hspace*{-2.2pt} [ #1 , #2\hspace*{1.5 pt} ] 
\hspace*{-2.7pt} \} }
\newcommand{\we}{\wedge}
\newcommand{\dv}{d^V}
\newcommand{\nbrpq}[2]{\nbr{\xxi{#1}{1}}{\xxi{#2}{2}}}
\newcommand{\lieni}[2]{$\pounds$${}_{\stackrel{#1}{X}_{#2}}$  }

\newcommand{\rbox}[2]{\raisebox{#1}{#2}}
\newcommand{\xx}[1]{\raisebox{1pt}{$\stackrel{#1}{X}$}}
\newcommand{\xxi}[2]{\raisebox{1pt}{$\stackrel{#1}{X}$$_{#2}$}}
\newcommand{\ff}[1]{\raisebox{1pt}{$\stackrel{#1}{F}$}}
\newcommand{\dd}[1]{\raisebox{1pt}{$\stackrel{#1}{D}$}}
\newcommand{\nbr}[2]{{\bf[}#1 , #2{\bf ]}}
\newcommand{\der}{\partial}
\newcommand{\oo}{$\Omega$}
\newcommand{\Om}{\Omega}
\newcommand{\om}{\omega}
\newcommand{\eps}{\epsilon}
\newcommand{\si}{\sigma}
\newcommand{\Lm}{\bigwedge^*}

\newcommand{\inn}{\hspace*{2pt}\raisebox{-1pt}{\rule{6pt}{.3pt}\hspace*
{0pt}\rule{.3pt}{8pt}\hspace*{3pt}}}
\newcommand{\sro}{Schr\"{o}dinger\ }
\newcommand{\bm}{\boldmath}
\newcommand{\vol}{\omega}
               \newcommand{\dvol}[1]{\der_{#1}\inn \vol}

\newcommand{\bd}{\mbox{\bf d}}
\newcommand{\bder}{\mbox{\bm $\der$}}
\newcommand{\bI}{\mbox{\bm $I$}}

\newcommand{\be}{\beta} 
\newcommand{\ga}{\gamma} 
\newcommand{\de}{\delta} 
\newcommand{\Ga}{\Gamma} 
\newcommand{\gmu}{\gamma^\mu}
\newcommand{\gnu}{\gamma^\nu}
\newcommand{\ka}{\varkappa}
\newcommand{\hka}{\hbar \varkappa}
\newcommand{\al}{\alpha}
\newcommand{\lapl}{\bigtriangleup}
\newcommand{\psib}{\overline{\psi}}
\newcommand{\Psib}{\overline{\Psi}}
\newcommand{\derts}{\stackrel{\leftrightarrow}{\der}}
\newcommand{\what}[1]{\widehat{#1}}

\newcommand{\bx}{{\bf x}}
\newcommand{\bk}{{\bf k}}
\newcommand{\bq}{{\bf q}}

\newcommand{\omk}{\omega_{\bf k}} 
\newcommand{\lpl}{\ell}
\newcommand{\zb}{\overline{z}} 

\newcommand{\BPsi}{{\bf \Psi}} 
\newcommand{\BH}{{\bf H}} 
\newcommand{\BS}{{\bf S}} 
\newcommand{\BN}{{\bf N}} 


\section{Introduction }

The precanonical approach to field quantization 
\cite{qs96,ikanat2,ikanat3,ik-grav,ik2,ik3,ik4,ik5,ik-gq} 
is based on a generalization of the Hamil\-tonian 
formulation from mechanics to field theory which 
requires no space-time decomposition and 
manages to incorporate the field dynamics into 
a finite dimensional covariant analogue 
of the phase space. The generalizations
of this kind have been long known in the calculus 
of variations \cite{dw,w35,rund,ka,vonrieth,lepage,lepage2,dedeck}.  
 Being based on a covariant analogue of the Legendre transformation 
and the Hamilton first order canonical equations 
but still  independent of the picture of fields 
as infinite dimensional systems evolving with time, 
these generalizations are in a sense 
intermediate between the Lagrangian 
and the standard canonical Hamiltonian formulation, 
hence the name ``precanonical.''  
The simplest and the basic for the present paper 
example 
is the 
so-called De Donder-Weyl (DW) formulation \cite{dw,w35,rund,ka}: 
given a Lagrangian density 
$L(y^a, y^a_\mu, x^\nu)$, a function of 
field variables $y^a$, their space-time derivatives 
(first jets) $y^a_\mu$,  
and   space-time 
variables $x^\mu$,  
one introduces new 
Ham\-il\-tonian-like variables:   
 $p_a^\mu:=\der L / \der y^a_\mu$ ({\em polymomenta})  
 \mbox{\rm and} 
  $H=H(y^a,p^\mu_a,x^\nu):=y^a_\mu p^\mu_a -L$    
 ({\em the DW Hamiltonian function}),  
and writes the field equations in the 
appealing  
Hamiltonian-like  first order form  \cite{dw,w35,rund,ka}
\beq
\der_\mu y^a (x) = \frac{\der H}{\der p^\mu_a}, 
\quad 
\der_\mu  p^\mu_a (x) =- \frac{\der H}{\der y^a }  ,  
\eeq 
provided the Lagrangian density is not singular 
in the sense that 
$\det \left (  ||\frac{\der^2 L}{\der y^a_\mu \der y^b_\nu}   ||\right ) 
\neq 0$. 

In this formulation fields are viewed as sections $y^a (x)$ 
in a  bundle of the field 
variables over the space time,  
a finite dimensional covariant configuration bundle 
with the local coordinates $(y^a,x^\mu)$.  
In a sense the formulation (1.1) can be considered as a 
``multi-temporal'' generalization of the Hamiltonian formulation 
to field theory since the space and time variables 
are present in (1.1) on equal footing as analogues 
of the time variable in mechanics. As a byproduct, 
this treatment makes the formulation applicable even 
when the space-time is not globally hyperbolic. The analogue of the 
phase space involved in (1.1) is a finite dimensional space of 
variables $(y^a, p_a^\mu, x^\mu)$ called the (extended) 
{polymomentum phase space}.

The manifest covariance and 
the 
``finite dimensionality'' in the 
aforementioned 
sense are the common features of a whole class of Hamiltonian-like 
formulations known from the theory of Lepagean equivalents in the 
calculus of variation \cite{lepage,lepage2,dedeck,krupka,krupka2,gotay-ext}.   
The above mentioned De Donder-Weyl formulation  represents the most 
straightforward covariant generalization of the Hamiltonian 
formulation in mechanics. Other Lepagean theories are based on more 
sophisticated definitions of polymomenta and the 
covariant Hamiltonian functions $H$ and still remain 
little known.  

The appropriate Poisson bracket operation for the 
precanonical DW 
formulation (1.1) has been introduced in 
\cite{ikanat,bial96,go96} 
(see also \cite{paufler,paufler2,roemer,helein} for recent 
attempts of generalizations). 
The Poisson brackets are defined on appropriate differential forms,   
which play 
the  role of dynamical variables, and lead to a Gerstenhaber 
algebra structure 
generalizing the Poisson algebra of observables 
in classical mechanics. The structures of the Hamiltonian formalism in 
mechanics are reproduced in the case of $(0+1)$-dimensional ``field 
theory''. The Poisson bracket on forms enables us to specify the notion 
of (pre)canonically conjugate field  and 
polymomentum 
variables and to write the 
DW Hamiltonian equations (1.1) 
in Poisson bracket formulation in which the 
bracket with $H$ is related to the 
operation of the total exterior differentiation, the latter generalizing 
the total time derivative in the Poisson bracket formulation of the 
equations of motion in mechanics 
(see \cite{ikanat2,ikanat,bial96,go96} 
for more details). 

Quantization based on these structures leads to the idea of a partial 
differential generalized 
Schr\"odinger equation for the wave {\em function} on the 
covariant configuration space: $\Psi=\Psi(y^a, x^\mu)$. The following 
Dirac-like equation has been put forward as a ``multi-temporal'' 
covariant generalization 
of the Schr\"odinger equation from quantum mechanics 
to field theory \cite{qs96,ikanat2,ikanat3}: 
\beq 
i\hbar\varkappa \gamma^\mu\der_\mu \Psi = \what{H}\Psi ,  
\eeq
where $\what{H}$ denotes the 
operator of the DW Hamiltonian, 
and 
$\varkappa$ is a (large) constant of the dimension 
 {\tt length}$^{-3}$  (in $n=4$ space-time dimensions) 
which 
 is interpreted as 
the ultraviolet cutoff of the volume of 
$\bk$-space 
(or 
an inverse characteristic ``minimal volume'' in $\bx$-space); 
and $\Psi=\Psi(y^a, x^\mu)$ 
is a Clifford-- (or spinor--) 
 valued wave function of a quantized field. 
Note that $\varkappa$ appears on purely dimensional grounds. 

The probabilistic interpretation of the wave function 
$\Psi(y^a, x^\mu)$ 
follows from the positive definiteness 
of \, $\Psib \gamma^0 \Psi$  \, 
for the Dirac spinor valued $\Psi$-s and from 
the conservation law 
\beq
\der_\mu \int \! dy \, \Psib \gamma^\mu \Psi = 0 
\eeq
which can be derived  from (1.2). 
One can interpret $\Psi(y^a, x^\mu)$ as 
the probability amplitude of finding the field value $y^a$ 
in (the vicinity of) the space-time point $x^\mu$. 
Obviously, the description of quantized fields in terms of this 
wave function is 
  fundamentally 
different from any standard description of quantum fields.   

The choice of the wave equation (1.2) is based 
on the observations that 
(i) it fulfils an analogue of the 
Ehrenfest theorem in the sense that the DW Hamiltonian equations 
(1.1) can be 
derived from (1.2) as the equations for the expectation values 
of properly chosen operators \cite{ikanat2,ikanat3},  
and that (ii) in the classical limit  
(1.2) can be reduced 
to the Hamilton-Jacobi 
equation of the De Donder-Weyl theory 
(see Sect. 3 and \cite{dw,w35,rund,ka,vonrieth}) 
by means of the ansatz 
$\Psi=Re^{iS^\mu \gamma_\mu/\hbar\varkappa}$ \cite{qs96,ikanat3}, 
where $S^\mu$ are 
the eikonal functions of the DW Hamilton-Jacobi theory (see Sect. 3). 

Quantization of a small Lie subalgebra of canonically conjugate 
field and polymomentum variables (given by proper differential forms)  
\cite{ikanat3,ikanat} 
suggests the following 
operator realization of polymomenta \cite{ikanat2,ikanat3} 
\beq
\hat{p}{}^\mu_a = - i\hbar\varkappa \gamma^\mu \frac{\der}{\der y^a}, 
\eeq
where the constant $\varkappa$ appears again on dimensional grounds. 

It is noteworthy that the ``multi-temporal'' 
quantum theoretic formalism which appears 
here as  a way of description of quantized fields    
 can be viewed as 
a hypercomplex extension of the formalism of 
quantum mechanics in which the wave functions and operators 
take values in the Clifford algebra of the underlying 
space-time manifold. The quantum mechanics 
appears then as a particular case 
corresponding to a $(0+1)$-dimensional field theory, when the 
corresponding Clifford algebra of $(0+1)$-dimensional space-time 
is just the algebra of the complex numbers. 

The elements of precanonical quantization as outlined above, 
of course, differ significantly from 
any presently known description of quantum fields. Their 
relation to the elements of the commonly practiced quantum field 
theory 
are far from being immediately obvious. This is the purpose of the 
present paper to clarify at least one aspect of this 
relation, namely an 
interplay between the wave function 
$\Psi (y^a, x^\mu)$ and the Schr\"odinger wave functional, and 
the interrelation between the Dirac-like wave equation (1.2) and 
the functional differential 
Schr\"odinger equation. The possibility of establishing of  
such a connection 
seems to be an 
important argument 
in favour of the wave equation (1.2) and precanonical quantization. 

In Section 2 we recall the elements of 
canonical and precanonical quantization 
of the scalar field which are used in what 
follows. An interplay between the functional differential Hamilton-Jacobi 
equation of the canonical formalism and the partial differential 
Hamilton-Jacobi equation of the precanonical De Donder-Weyl theory 
is discussed in Section 3. This result is extended to the quantum level 
in Section 4 where we show how the 
functional differential 
Sch\"odinger equation and the Sch\"odinger wave functional 
are derived from the partial differential  Dirac-like 
generalized Sch\"odinger equation of the precanonical approach 
and the corresponding Clifford-valued wave function. 
Concluding remarks are presented in Section 5.

\section{
Two Quantizations of the Scalar Field}   

In this section we outline canonical and precanonical quantization 
of the free scalar field given by 
the Lagrangian density 
(henceforth we set $\hbar = 1$) 
\beq 
L= \half \der_\mu y \der^\mu y - \half m^2 y^2 .  
\eeq 
Our main purpose is to 
present a background on both approaches necessary for the 
discussion of 
an interplay between them in the 
subsequent sections.

\subsection{Canonical quantization} 

 Let us recall some basic elements of canonical quantization 
of the scalar field in the functional Schr\"odinger 
representation  \cite{hatfield,schw,mansf}. 
The canonical momentum is defined as 
$\pi (\bx) = \der L/ \der (\der_t y(\bx))$ and the 
canonical Hamiltonian functional, 
which is a functional of $y (\bx)$ and $\pi (\bx)$,  
as 
$\BH := \int \! d\bx \, (\der_t y (\bx)\pi (\bx) -L  ).$\footnote{Here 
and in what 
follows the bold capital letters are 
reserved 
to designate functionals; 
$\bx=x^i$ denote the 
spatial components of a four-vector $x^\mu := (x^i,x^t)$. }  
From the Lagrangian (2.1) one obtains: 
$\pi (\bx) = \der_t y(\bx)$ and 
\beq 
\BH = \half \int \! d\bx \, \left ( \pi^2(\bx) + (\der_i y(\bx))^2 
+ m^2 y^2(\bx)\right ). 
\eeq 
Quantization of the canonical Poisson brackets leads to the 
realization 
\beq
\hat{\pi}(\bx)=-i\frac{\delta}{\delta y(\bx)}
\eeq 
in the functional Scr\"odinger representation when the 
quantum states are described by a time dependent 
functional of field configurations $y(\bx)$: 
$\BPsi_S=\BPsi_S([y(\bx)],t)$ 
which fulfils the  functional Schr\"odinger equation 
\beq
i\der_t \BPsi_S = \what{\BH}_S \BPsi_S , 
\eeq 
where the functional Hamiltonian operator 
for the scalar field has the form 
\beq
\what{\BH}_S = \half \int \! d\bx \, \left ( 
-\frac{\delta^2}{\delta y^2(\bx)}  + (\der_i y(\bx))^2  
+ m^2 y^2(\bx)\right ). 
\eeq
The ground (vacuum) state wave functional of this  Hamiltonian operator 
can be 
 represented in terms of the Fourier components of 
$y(\bx) = \int \! \frac{d\bk}{(2\pi)^{3}} \, y(\bk) e^{i\bk\bx}$ 
as follows:  
\beq
\BPsi_{S {\rm vac}}= \exp \left ( 
-\frac{1}{2} \int \! \frac{d\bk}{(2\pi)^{3}} \, 
\omega_\bk y(-\bk) y(\bk) \right )
\eeq
and it corresponds to the divergent energy eigenvalue 
\beq    
E_{\rm vac}=\half \int \! d\bx \, 
\int \! \frac{d\bk}{(2\pi)^{3}}\, \omega_\bk . 
\eeq     
The excited states are known to correspond to the 
multi-particle (Fock) states which can be generated 
by 
the 
iterated action of the creation operator  
\beq    
a^\dagger (\bk) = \frac{1}{\sqrt{2}}
\int \! d\bx \, e^{-i\bk\bx} 
\left ( -\frac{\delta}{\delta y(\bx)}  
+ \omega_\bk y(\bx)  \right ) 
\eeq   
on the vacuum state wave functional (2.6).

\subsection{Precanonical quantization} 

Now, let us quantize the scalar field theory 
using the 
precanonical 
procedure outlined in the Intro\-duct\-ion.      
The Lagrangian density (2.1) gives rise to the 
polymomenta $p^\mu = \der^\mu y$ 
and the DW Hamiltonian function 
\beq
H (y, p^\mu) =\half p^\mu p_\mu + \half m^2 y^2 . 
\eeq 
Precanonical quantizaton leads to the realization of the 
operators of polymomenta in the form 
\beq
\hat{p}{}^\mu = - i\varkappa \gamma^\mu \der_y ,  
\eeq 
as in (1.4), 
and to the following operator of the DW Hamiltonian 
\cite{qs96,ikanat2,ikanat3} 
\beq
\what{H} := -\half \varkappa^2 \der_{yy}^2 
+ \half m^2 y^2  . 
\eeq
In our previous papers \cite{qs96,ikanat3} we presented a solution of the 
generalized Schr\"odinger equation (1.2) with the DW 
Hamiltonian operator (2.11) in the basis where the latter is diagonal. 
However, the fact that there is no physical interpretation of 
$H$ on the classical level implies that this basis is not physical. 
One should diagonalize an operator representing a physical quantity, 
such as the energy. 

Obviously, one can rewrite (1.2) 
in the space-time decomposed form 
\beq
i\der_t \Psi = -i \alpha^i\der_i \Psi  + 
\beta \frac{1}{\varkappa}\what{H}\Psi =: \what{{\cal E}} \Psi 
\eeq 
and interpret the operator $\what{{\cal E}}$ defined 
in the right hand side 
as the energy operator,  
as it represents the time evolution. 

Equation (2.12) is best solved in the Fourier space where it takes 
the form
\beq 
k^t \Psi (y, \bk) = \left ( 
- \alpha^i \bk_i
+ \beta \frac{1}{\varkappa}\what{H} \right ) \Psi (y, \bk) , 
\eeq 
where $k^\mu := (\bk^i, k^t)$. 
Taking into account that $m^2$ in (2.11) can be written as 
$m^2=:(q_\mu\gamma^\mu)^2$ for some $q^\mu = (\bq, \omega_\bq)$, 
$\omega_\bq:=\sqrt{m^2 + \bq^2}$,   
one can 
find the ground state $(N=0)$ 
solution 
(up to a normalisation factor) 
\beq
\Psi_0 (y, \bq) = e^{-\frac{1}{2\varkappa} q_\mu\gamma^\mu y^2 }, 
\eeq 
which corresponds to the eigenvalues $k^t_0=\half \omega_\bq$, 
$k^i_0=\half q^i$. 
The higher excited states can be 
easily found to correspond to $k^\mu_N = (N+\half) q^\mu$. 
They are  generated by acting 
by the creation operator 
\beq
a^\dagger (\bq)=\frac{1}{\sqrt{2}} \left (-\sqrt{\varkappa} \der_y 
+ \frac{1}{\sqrt{\varkappa}}q_\mu \gamma^\mu y \right )  
\eeq 
on the ground state (2.14).

\section{Canonical vs. Precanonical I: Hamilton-Jacobi equations} 

A brief comparison of the above two approaches to field quantization 
seems to reveal a huge conceptual distance between 
the both: whereas one is based on the language of functionals and 
functional differential equations the second one uses only 
(Clifford-valued) functions and partial differential equations.   
In view of the well established success of the canonical quantization approach 
this might mean a no-go sentence to the precanonical approach.
This conclusion, however, appears to be too hasty. In this section 
we demonstrate 
how 
both approaches can be related to each other 
on the classical level of Hamilton-Jacobi theory. This consideration 
hints to the existence of a similar 
relation 
between precanonical 
and canonical approach on the quantum level and 
partially suggests a 
technique of establishing such 
a relation.  

Various aspects of the 
interplay between the precanonical 
DW theory and the canonical Hamiltonian formalism 
 have been discussed earlier 
\cite{ikanat,helein,gimmsy,regge,gotay1,gotay2,sniat}. 
It has been shown that a general idea of relating the 
precanonical quantities and structures to the canonical ones 
is (i) to perform the space-time decomposition, 
(ii) to restrict the precanonical quantities to the
Cauchy surface $\Sigma$ in the covariant configuration 
space, $\Sigma$: $( y = y(\bx), t={\rm const})$, 
and then (iii) integrate over it. 

 Here we will show how the Hamilton-Jacobi equation of the 
De Donder-Weyl theory (DWHJ equation) 
\cite{rund,ka,vonrieth} 
\beq
\der_\mu S^\mu + H \left 
(y^a, p_a^\mu = \frac{\der S^\mu}{\der y^a }\right ) = 0
\eeq
can be related to the 
functional differential Hamilton-Jacobi equation of the 
canonical formalism: 
\beq
\der_t \BS + \BH \left ( 
y(\bx), \pi (\bx) = \frac{\delta \BS}{\delta y(\bx)} \right ) 
=0.   
\eeq  
We basically shall argue that the following functional constructed 
from the DW eikonal functions $S^\mu (y^a, x^\mu)$ 
\beq 
\BS := \int_\Sigma \left (S^\mu \omega_\mu \right ) |_\Sigma = 
\int_\Sigma S^t|_\Sigma 
\, d\bx , 
\eeq
where $\omega_\mu:= \der_\mu \inn (dx^1\we dx^2\we dx^3\we dx^t)$  
and $S^t|_\Sigma := S^t (y^a=y^a(\bx), \bx, t)$ is a restriction of 
the time-like component of $S^\mu (y^a, x^\mu)$'s to $\Sigma$,  
fulfills the standard functional differential HJ equation as a 
consequence of the 
partial differential DWHJ equation for $S^\mu$'s. 

For simplicity, let us consider the specific case of 
the scalar field. Then the DWHJ equation takes the form 
(see (2.8) and (3.1)) 
\beq
\der_t S^t + \der_i S^i + \half \der_y S^\mu \der_y S_\mu + 
\half m^2 y^2 = 0 
\eeq 
and the standard functional differential HJ equation reads 
(see (2.2) and (3.2))
\beq
\der_t \BS + \half \int \! d\bx \, \left ( 
\left ( \frac{\delta \BS}{\delta y(\bx)} \right )^2 + 
\left (\der_i y(\bx)\right )^2 + m^2 y(\bx)^2 \right ) =0 . 
\eeq
From (3.3) we obtain 
\beqa 
\der_t \BS &=& \int \! d\bx \, \der_t S^t|_\Sigma , \\
\frac{\delta \BS}{\delta y(\bx)} &=& \der_y S^t|_\Sigma . 
\eeqa
The equation for $S^\mu|_\Sigma$ can be obtained from 
the DWHJ equation 
by noticing that when acting on $S^\mu|_\Sigma$ 
the spatial derivative $\der_i$ turns into the total 
derivative $\frac{d}{dx^i} := \der_i + \der_i y(\bx)\der_y$ 
the last term of which should be compensated.  
Therefore, the equation for $S^\mu|_\Sigma$ assumes the 
form 
(in the metric signature $- - - +$)  
\beq
\der_t S^t|_\Sigma + \frac{d}{dx^i} S^i|_\Sigma
- \der_i y(\bx) \der_y S^i|_\Sigma 
+\half \left ( \der_y S^t|_\Sigma \right )^2 - 
\half \left (\der_y S^i|_\Sigma \right )^2 
+ \half m^2 y(\bx)^2 = 0 . 
\eeq 
Substituting $\der_t S^t|_\Sigma $ from this equation 
into the right hand side of (3.6) and using (3.7) we obtain 
$$ 
\der_t \BS +  \int \! d\bx \, \left ( 
\half\left (  \frac{\delta \BS}{\delta y(\bx)} \right )^2 
+ \frac{d}{dx^i} S^i|_\Sigma
- \der_i y(\bx) \der_y S^i|_\Sigma  
 - \half \left (\der_y S^i|_\Sigma \right )^2  
+\half  m^2 y(\bx)^2 \right ) =0 . 
$$ 
The second term under the integral does not contribute 
being a total divergence. The third and the forth terms 
together lead to $\half (\der_i y(\bx))^2$ because 
in the DWHJ theory $\der_y S^i = p^i$ 
and, for the scalar field, $p^i|_\Sigma =- \der_i y(\bx)$. 
We, therefore, obtained the functional differential 
HJ equation, eq.~(3.5), 
as a consequence of  
the DWHJ equation (3.4) 
restricted to the Cauchy surface $\Sigma$ 
and a natural 
hypothesis (3.3) as to how the Hamilton-Jacobi 
eikonal functional $\BS$ is 
related to the DWHJ eikonal functions $S^\mu$. 


Similar idea of the restriction of a precanonical equation 
to the Cauchy surface $\Sigma$ will be used below to 
clarify  the interrelation  
between the functional Schr\"odinger equation and our proposed 
Dirac-like 
generalized Schr\"odinger equation for quantized fields. 
Since  the Hamilton-Jacobi equations (3.4) and (3.5) 
correspond to the classical limit 
of the Schr\"odinger equations (1.2), (2.11) 
and  (2.4), (2.5), respectively,  
this  interrelation appears now to be quite natural.

\section{Canonical vs. Precanonical II: wave function(al)s and 
wave equations}

As we have mentioned above,  
the wave function $\Psi (y^a, x^\mu)$ in eq. (1.2) 
is 
 the 
probability amplitude of finding the values  $y^a$ 
of the  fields  
in the space-time point $x^\mu$. It is natural to ask 
how this quantity could be related to the 
Schr\"odinger wave functional $\BPsi_S ([y^a(\bx)], t)$ 
which is 
 known to be 
the 
probability amplitude of observing the field 
{\em configuration} $y^a(\bx)$ on a space-like hypersurface 
of constant time $t$. A related question is 
how the functional differential 
equation for the Schr\"odinger wave functional $\BPsi_S ([y^a(\bx)], t)$ 
is related to the Dirac-like partial differential equation 
for the function $\Psi (y^a, x^\mu)$. 

An obvious idea is to view the probability amplitude of observing 
a configuration $y^a(\bx)$  as a composition of single amplitudes of 
observing the corresponging values of the field 
$y^a=y^a(\bx)$ at each point $\bx$.   
For this purpose we restrict the wave function $\Psi (y^a, x^\mu)$ 
to a Cauchy surface $\Sigma$: $t=const, y^a=y^a(\bx)$, 
to obtain 
the function $\Psi|_\Sigma:=\Psi (y^a(\bx), \bx, t)$ which is a 
probability amplitude of observing the specific value 
$y^a(\bx)$ of 
the field in the given 
spatial point $\bx$ (at the moment of time $t$),   
 and then construct 
a {\em joint} probability amplitude of 
simultaneously observing the respective 
values $y^a=y{}^a(\bx)$ 
at all points $\bx$ 
of an equal-time hypersurface with the time label $t$. 
This amplitude, 
to be denoted $\BPsi ([y^a(\bx)], t)$,  
then describes the same as 
the Schr\"odinger wave functional does. 

If there are no correlations 
 between the amplitudes 
 $\Psi (y^a(\bx), \bx, t)$ 
at space-like separated points 
the composed amplitude  $\BPsi ([y^a(\bx)], t)$ 
can be represented as the product 
of the single amplitudes given by 
 $\Psi|_\Sigma$ 
over all 
spatial points $\bx$, i.e.  
\beq
 \BPsi ([y^a(\bx)], t) = \prod_{\bx\in\Sigma} \Psi|_\Sigma 
= \prod_\bx \Psi (y^a(\bx), \bx, t) .  
\eeq 
The above continual product expression can be 
interpreted as follows: 
\beq
\prod_\bx \Psi (y^a(\bx), \bx, t)  
= e^{\sum_\bx \ln \Psi (y^a(\bx), \bx, t) }  
= \lim_{\Delta x \rightarrow 0} 
e^{\frac{1}{(\Delta x)^3} \int \! d\bx \, \ln \Psi (y^a(\bx), \bx, t)},  
\eeq 
if 
the ordering problem of the matrix valued 
$\Psi|_\Sigma$'s at different points $\bx$  
and the problem of the existence of the limit 
are ignored.  This symbolic formula motivates 
the following well-defined expression for the wave functional 
which is a quintessence of the above physical idea:  
\beq
\BPsi ([y^a(\bx)], t) = 
  {\rm tr} \left ( 
(1+\beta) e^{ \varkappa \int \! d\bx \, \ln \Psi (y^a(\bx), \bx, t)}  
\right ) . 
\eeq 
In (4.3)  the trace serves to produce a scalar functional 
from the matrix valued one in (4.2), 
the (``very large'') constant $\varkappa$, according to its 
physical meaning of the ultraviolet cutoff in the $\bk$-space, 
replaces  $1/(\Delta x)^3$ in (4.2), and the 
projector to the  posititive  
frequency part, $(1+\beta)$, is included under the trace 
for the reasons which will be shortly clarified. 

Let us find now a functional differential 
equation satisfied by the functional (4.3) as 
a consequence of equation 
(2.12) 
for the wave function $\Psi (y^a, x^\mu)$ 
or, more precisely, of the corresponding equation 
for  its restriction to the Cauchy surface $\Sigma$, 
$\Psi|_\Sigma := \Psi (y^a(\bx), \bx, t)$.  
This equation is derived from (2.12) by the 
procedure which is identical to the one of 
deriving 
of equation (3.8) for $S^\mu|_\Sigma$ from the DWHJ equation for $S^\mu$. 
It yields:   
\beq
i\der_t \Psi|_\Sigma = -i\al^i \frac{d}{dx^i} \Psi|_\Sigma 
+ i\al^i\der_i y^a (\bx)\frac{\der}{\der y^a} \Psi|_\Sigma 
+  \frac{1}{\varkappa} \beta \ \what{H} \Psi|_\Sigma . 
\eeq 

Now, for the time derivative of the functional 
$\BPsi ([y(\bx)], t) $ 
given by (4.3) we obtain  
\beq
i\der_t \BPsi ([y(\bx)], t) 
= {\rm tr} \left \{ (1+\beta) e^{\varkappa \int \! d\bx \, \ln \Psi}  
\varkappa \int \! d\bx \, \Psi^{-1} i\der_t \Psi
 \right \},  
\eeq 
where here and throughout the following calculation $\Psi$ denotes 
$\Psi|_\Sigma = \Psi (y(\bx), \bx, t)$   
and the consideration is confined, 
for simplicity,  
to the case of the one-component scalar field $y$. 
In what follows we also assume that $\Psi$ 
is an invertible matrix, though this assumption can be avoided 
by setting $\Psi=e^W$ and working dirctly 
in terms of $W$ instead of $\ln \Psi$. 

Further, 
\beq
\frac{\delta\BPsi}{\delta y(\bx)} = 
{\rm tr} \left \{ (1+\beta) e^{\varkappa \int \! d\bx \, \ln \Psi} 
\varkappa \Psi^{-1}  \der_y \Psi 
 \right \},  
\eeq 
and 
\beqa
\frac{\delta^2\BPsi}{\delta y(\bx)^2} 
&=&  
{\rm tr} \left \{ (1+\beta) e^{\varkappa \int \! d\bx \, \ln \Psi} 
\left ( \varkappa^2 \Psi^{-1}  \der_y \Psi \, \Psi^{-1}  \der_y \Psi 
- \varkappa \delta (0) \Psi^{-1}\der_y \Psi \Psi^{-1} \, \der_y \Psi 
\right. \right.\nn \\
&&
\left. \left. 
+ \varkappa \delta (0)\Psi^{-1} \der^2_{yy} \Psi 
\right ) 
\right \} . 
\eeqa 
As usual, the second functional derivative 
at equal points is not well-defined   
because of the terms involving $\delta(\bx=0)$ present.   
It requires a regularization which essentially 
 amounts to 
replacing $\delta(0)$ with 
 the 
momentum space cutoff. However, the latter  
already has its counterpart 
in our theory 
as the constant $\varkappa$. 
Under the regularization which identifies 
$\delta(0)$ with $\varkappa$ 
the first two terms 
in (4.7) cancel each other and we obtain
\beq
\frac{\delta^2\BPsi}{\delta y(\bx)^2} 
= {\rm tr} \left \{ (1+\beta) e^{\varkappa\int \! d\bx \, \ln \Psi} 
\varkappa^2 \Psi^{-1} \der^2_{yy} \Psi 
\right \} . 
\eeq

Now, 
the substitution of (4.4) into the right hand side of (4.5) 
yields 
$$ 
i\der_t \BPsi 
 = {\rm tr} \left \{ (1+\beta) e^{\varkappa \int \! d\bx \, \ln \Psi}  
\int \! d\bx \,  
\Psi^{-1} 
 \left ( -i\varkappa \alpha^i\frac{d}{dx^i} \Psi 
 + i\varkappa \al^i\der_i y (\bx)\der_y \Psi 
+ \beta \what{H} \Psi 
\right )  \right \}.   
$$ 
The first term under the integral reduces to the total 
divergence $\frac{d}{dx^i}(\alpha^i \ln \Psi)$ 
and does not contribute.  
The remaining terms lead to 
\beq 
i\der_t \BPsi 
= {\rm tr} \left \{ (1+\beta) e^{\varkappa\! \int \! d\bx \, \ln \Psi}  
\int \! d\bx \,  
\Psi^{-1}
 \left ( i\varkappa \gamma^i\der_i y (\bx) \der_y \Psi 
+ \what{H} \Psi 
\right )  \right \},  
\eeq 
where it is used that 
$(1+\beta)\beta = (1+\beta) $, $\beta^2 = 1$.  
This is where the projection to the 
positive frequency part in (4.3) is 
crucial for the argument.  

Substituting (2.11) to (4.9) and comparing the result with 
(4.8) we obtain 
\beqa
i\der_t \BPsi ([y(\bx)], t) &=&  \half \int d\bx \left 
( -\frac{\delta^2}{\delta^2 y (\bx)} 
+ m^2 y^2(\bx) 
\right ) \BPsi ([y(\bx)], t) 
\nn \\ 
&& +  
 {\rm tr} \left \{(1+\beta) e^{ \varkappa \! \int \! d\bx' \, \ln \Psi} \,  
\int \! d\bx \,  i \varkappa \der_i y (\bx) \gamma^i \Psi^{-1} \der_y \Psi 
\right \} . 
\eeqa
The first two terms in (4.10) are identical to the corresponding 
terms in the functional differential 
Schr\"odinger equation, eqs. (2.4), (2.5).  
Let us turn to the last term which,   
with the aid of (4.6),  
can be written in the form 
\beq 
 {\rm tr}\left \{  \int \! d\bx \,  \der_i y (\bx)
i \gamma^i \frac{\delta}{\delta y(\bx) }
||\BPsi || 
\right \} ,  
\eeq  
where $||\BPsi ||$ is the matrix-valued functional given by 
\beq
||\BPsi || := 
(1+\beta) e^{ \varkappa \int \! d\bx \, \ln \Psi (y(\bx), \bx, t)} , 
\eeq
such that $\BPsi = {\rm tr}(||\BPsi ||)$.  

In order to interpret this term 
let us notice  that the canonical Hamiltonian density 
${\cal H} = T^t_t:= p^t(\bx) \der_t y (\bx) - 
L $ 
can be represented in the form 
(c. f. the definition of $H$ in (1.1))  
\beq 
{\cal H} (\bx) = H |_\Sigma (\bx) - \der_i y (\bx) p^i|_\Sigma (\bx) , 
\eeq 
where $H|_\Sigma  (\bx)$ and $p^i |_\Sigma (\bx)$ denote the 
quantities of the DW theory, the DW Hamiltonian $H$ and 
the spatial polymomenta $p^i$, restricted to 
the Cauchy surface $\Sigma$. We know that the operators  
of space-like polymomenta $p^i$ in the precanonical scheme 
are given by 
 $- i \varkappa \gamma^i \der_y$ (c. f. eq. (2.10)). 
Then (4.6) suggests that 
 their restriction to $\Sigma$, 
the operators  of $p^i (\bx)$, act on the functionals (4.3) 
 as follows: 
\beq
{\rm tr} \left \{ \hat{p}{}^i(\bx) 
||\BPsi || 
\right \} 
= - {\rm tr} \left \{
i \gamma^i \frac{\delta}{\delta y (\bx)}  
 ||\BPsi || 
\right \} . 
\eeq
Therefore, the expression in (4.11) 
is naturally identified with 
\beq
- 
{\rm tr} \left \{  \int \! d\bx \, 
\der_i y (\bx) 
\hat{p}{}^i(\bx) 
||\BPsi || 
\right \} , 
\eeq 
where the second term in (4.13) can be recognized. 
Consequently, eq. (4.10) can be viewed as the trace 
of the matrix functional differential 
Schr\"odinger equation 
\beq 
i\der_t ||\BPsi|| - \what{{\BH}}\, ||\BPsi||   
= 0 , 
\eeq
where  $\what{{\BH}} $ is the matrix-valued 
 functional differential operator 
 corresponding to the Hamiltonian density written in the form (4.13):  
\beq
\what{{\BH}} = \int \! d\bx \,
 \what{{\cal H}} (\bx) 
 = \int \! d\bx \, \left 
( - 
 \half  
\frac{\delta^2}{\delta^2 y (\bx)} 
+ 
 \half  
m^2 y^2(\bx)
+ 
i \der_i y(\bx) \gamma^i \frac{\delta}{\delta y (\bx)} 
\right ) . 
\eeq 

Obviously, one would like to understand in greater details 
a relation between 
the Hamiltonian operator in (4.17) 
and the standard Schr\"odin\-ger picture Hamiltonian operator 
$\what{{\BH}}_S$, eq. (2.5).  
One can immediately ask if there exists a 
unitary transformation which relates (4.17) to (2.5):  
\beq
\what{{\BH}}_S = e^{-i\BN}\what{{\BH}}e^{i\BN} , 
\eeq 
where $\BN$ is a functional operator to be found. 
A straightforward calculation yields 
\beqa
e^{-i\BN}\what{{\BH}}e^{i\BN}
&=&  \half \int \! d\bx \, 
\left (-\frac{\delta^2 i\BN }{\delta^2 y (\bx)} 
- \left ( \frac{\delta i\BN}{\delta y (\bx)} \right )^ 2 
- 2 \frac{\delta i\BN}{\delta y (\bx)}\frac{\delta }{\delta y (\bx)} 
- \frac{\delta^2  }{\delta^2 y (\bx)} 
\right.
\nn \\ 
&& \left.
\hspace*{-20pt} 
+ \; 
2i e^{-i\BN} \der_i y(\bx) \gamma^i e^{i\BN} 
 \left ( \frac{\delta i\BN}{\delta y (\bx)} 
+ \frac{\delta }{\delta y (\bx)}\right ) 
+ m^2 e^{-i\BN}{y^2(\bx)}e^{i\BN}  
\right ) . 
\eeqa 
In order to obtain ${\BH}_S$ in the right hand side of (4.19),  
we have to ensure first that the terms proportional to 
$\frac{\delta }{\delta y (\bx)} $ compensate each other. 
This requirement leads to the equation 
\beq
\frac{\delta \BN }{\delta y (\bx)} =  \gamma^i \der_i y (\bx) . 
\eeq 
Then 
$\frac{\delta^2 \BN }{\delta^2 y (\bx)} = \gamma^i\der_i \delta (0) = 0 $ 
and  
the other two terms in (4.19) yield 
\beqa 
\left ( \frac{\delta i\BN}{\delta y (\bx)} \right )^ 2 
&=&  (\der_i y (\bx))^2 ,  
\nn \\
2i e^{-i\BN} \der_i y(\bx) \gamma^i e^{i\BN} 
 \frac{\delta i\BN}{\delta y (\bx)} 
&=& 2 (\der_i y (\bx))^2 , 
\eeqa  
which are consistent with each other and with (4.20). 
The solution to (4.20) can be written in the form 
\beq
\BN=\BN [y(\bx), \Sigma] =  
 \int_\Sigma  y (\bx) \gamma^i d \Sigma_i ,  
\eeq
where $d \Sigma_i:= dy\we \der_i \inn (dx^1\we dx^2\we dx^3)$ 
(note, that $\Sigma$ is not varied under $\frac{\delta}{\delta y (\bx)}$). 
Obviously, the unitary tranformation generated by $\BN$ 
leaves the $m^2 y^2(\bx)$ term in (4.17) invariant. 
Hence, 
summing up the results we obtain  
\beq
e^{-i\BN} \what{{\BH}} e^{i\BN} = \half \int \! d\bx \, 
\left (- \frac{\delta^2  }{\delta^2 y (\bx)} 
+ (\der_i y (\bx))^2 
+ m^2 y^2(\bx) \right ) 
= \what{{\BH}}_S . 
\eeq 

Thus, it is proven that the matrix-valued Hamiltonian operator 
(4.17) can be unitarily transformed to the standard 
Hamiltonian operator in the standard Schr\"odinger 
representation, eq. (2.5). 
Correspondingly, from (4.10) and (4.16) it follows  that 
the wave functional 
\beq
\BPsi_S = {\rm tr} \left ( 
e^{-i\BN} ||\BPsi ||  \right )  
\eeq 
 fulfills the standard functional differential 
 Schr\"odinger equation, eqs. (2.4) and (2.5). 
 It is, therefore, the wave functional in the 
 Schr\"odinger representation. 
Let us stress that the present results follow from 
the formula (4.3) constructing a wave functional 
$\BPsi ([y(\bx)],t)$ from the Clifford-valued 
wave function $\Psi (y^a, x^\mu)$ 
and the Dirac-like 
covariant generalized Schr\"odinger 
equation for the latter.

\subsection{An example: the vacuum state wave functional}

Let us demonstrate how the formula (4.3) 
can be used to construct the vacuum state wave 
functional, eq. (2.6), from the ground state wave function 
of the precanonical approach, eq. (2.14). The latter 
yields $\ln \Psi_0 (y,\bq) = -\frac{1}{2\varkappa}q^\mu\gamma_\mu y^2$.   
The restriction 
$(\ln \Psi_0)|_\Sigma (\bx)$ in constructed 
by taking into account that 
$$y|_\Sigma (\bx) = y(\bx) = 
\int \! \frac{d\bq}{(2\pi)^{3}} \, e^{i\bq\bx} y(\bq)$$  
and assuming that 
\beq
\left (q^\mu\gamma_\mu y \right )|_\Sigma (\bx) 
:= \int \! \frac{d\bq}{(2\pi)^{3}} \, e^{i\bq\bx} q^\mu\gamma_\mu y(\bq). 
\eeq 
This allows us to write 
\beqa
\int \! d\bx \, \left (\ln \Psi_0 \right ) |_\Sigma 
&=& -\frac{1}{2\varkappa} \int \! d\bx \, 
\int \! \frac{d\bq}{(2\pi)^{3}} \, e^{i\bq\bx}  
q^\mu\gamma_\mu y(\bq) 
\int \! \frac{d\bq{}'}{(2\pi)^{3}} \, e^{i\bq{}'\bx} y(\bq') 
\nn \\ 
&=& -\frac{1}{2\varkappa} \int \! \frac{d\bq}{(2\pi)^{3}} \, 
\beta \omega_\bq y(-\bq) y(\bq) . 
\eeqa 
Substituting this result to (4.3) we obtain the standard vacuum state 
wave functional for the quantum scalar field, eq. (2.6): 
\beq
{\BPsi}_{\rm vac} = \exp \left ( 
-\frac{1}{2} \int \! \frac{d\bq}{(2\pi)^{3}} \, 
\omega_\bq y(-\bq) y(\bq) \right ) ,  
\eeq 
as a result of the specific composition of the Clifford algebra 
valued ground state wave functions (2.14) 
as given by the formula 
(4.3).

\section{Conclusions} 

In this paper we have demonstrated how the  standard 
functional Schr\"odinger representation in quantum 
field theory can be derived from the recently proposed 
precanonical field quantization which is based 
on a finite dimensional manifestly covariant  
generalization of the Hamiltonian formalism to field theory 
and leads to 
a 
covariant 
Dirac-like ``multi-temporal'' generalization of 
the Schr\"odinger equation on a finite dimensional configuration 
bundle of field variables over the space-time. 
Note that those are 
sections of this bundle which are the field configurations 
whose space is infinite dimensional and underlies the conventional 
canonical quantization. 

We argued that 
the Schr\"odinger wave functional appears to be 
related to the continual product, over all spatial points, of 
the values of the wave function fulfilling the above Dirac-like 
equation restricted to a Cauchy data surface.\footnote{
 An earlier discussion of a possible 
continual product structure of the Schr\"odinger wave functional 
 can be found in \cite{qs96,ikanat3,tapia}  
(c. f. \cite{navarro} for a related brief discussion but 
a conclusion different from ours). 
Such a structure is also characteristic to the ultralocal models 
extensively studied by 
Klauder \cite{klauder}.}  
In this sense the information contained in the Schr\"odinger 
wave functional can be extracted from the abovementioned 
wave function.  In doing so the constant 
$\varkappa$ appearing in precanonical quantization 
is interpreted as a quantity of the ultraviolet cutoff scale.

The  existence of the connection 
between the canonical and precanonical 
approaches to field quantization is an additional argument in favor 
of precanonical quantization, 
 even though 
the latter is yet to be developed to the level of 
a calculational technique in quantum field theory. 
This connection is not restricted to the simple case of a free 
scalar field treated here. In fact, it can be 
straightforwadly extended to the interacting scalar fields. 
Moreover, a preliminary consideration shows that it is also possible 
to extend the argument to other fields, such as 
the Yang-Mills 
or the spinor fields. 
However, the technical difficulty we face then is that those 
fields are singular with respect to the De Donder--Weyl Legendre 
transformation $y^a_\mu \rightarrow p^\mu_a, \; L\rightarrow H$ {}  
and, therefore, they require a properly modified 
precanonical quantization procedure, similarly to the 
quantization of  constrained systems. This work is in progress. 

Note that the common infinities of the canonical quantization 
approach, such as the vacuum energy or the singularity in the second 
functional derivative in the Hamiltonian operator, 
are not present in the precanonical approach. 
Our discussion clearly indicates that these infinities appear only 
when the functional objects are constructed from  the 
well defined quantities of the precanonical approach. 
However, 
not all infinities in quantum field theory are 
 likely to be of this nature 
because 
the characteristic singularities of Green functions are persistent  
also in the precanonical approach.

Note also that 
the equations of the 
precanonical approach remain valid in the space-times which do not 
admit the  
space-time decomposition, i.e. are not globally hyperbolic,  
and, therefore, do not enable us to introduce 
the functional Schr\"odinger representation in a straightforward way. 
In this sense  precanonical quantization seems to have a wider 
range of validity than canonical quantization,  
though its physical interpretation 
beyond the framework of globally hyperbolic space-times 
may require new physical insights. 

\section*{Acknowledgements} 
 I am grateful to E. Gozzi,  
M. C. Mu\~nos-Lecanda,  
N. Roman-Roy, 
H. R\"omer, 
and F. H\'elein 
for their kind hospitality during my visits to their departments 
which were stimulating for the present work. 
I also thank J. Klauder, for enlightening conversations on ulralocal models 
and for making the book \cite{klauder} available to me prior to publication, 
and V. Tapia, for useful discussion and for bringing 
Ref. \cite{tapia} to my attention. 
I am indebted to Prof. A. Wipf and the Institute of Theoretical 
Physics in Jena for their kind hospitality and the opportunity 
to pursue this research.

\end{document}